\documentclass[%
 reprint,
 amsmath,amssymb,
 aps,
]{revtex4-2}
\usepackage{xcolor}
\usepackage{dcolumn}
\usepackage{bm}
\usepackage{amsmath}
\usepackage{physics}
\usepackage{braket}
\usepackage{float}
\usepackage{graphicx}
\usepackage[position=top]{subfig}
\usepackage{subfloat}
\usepackage[normalem]{ulem}

\renewcommand{\eqref}[1]{Eq.~{(\ref{#1})}}
\newcommand{\figref}[1]{Fig.~\ref{#1}}
\DeclareMathOperator{\sinc}{sinc}

\usepackage[toc,page]{appendix}

\usepackage[linkcolor=blue,colorlinks=true]{hyperref}
\usepackage{cleveref}
\crefformat{equation}{#2equation~#1#3}
\Crefformat{equation}{#2Equation~#1#3}

\usepackage[T1]{fontenc}

\begin{document}

\preprint{APS/123-QED}
\title{High-Purity OAM-Entangled Photons from SPDC with Reduced Spatial-Spectral Correlations}

\author{F. Crislane V. de Brito}
\email{crislane.brito@umk.pl}
\affiliation{Institute of Physics, Faculty of Physics, Astronomy and Informatics, Nicolaus Copernicus University in Toruń, ul. Grudziądzka 5, 87-100 Toruń, Poland}

\author{Sylwia M. Kolenderska}
\affiliation{Institute of Physics, Faculty of Physics, Astronomy and Informatics, Nicolaus Copernicus University in Toruń, ul. Grudziądzka 5, 87-100 Toruń, Poland}
\affiliation{School of Physical and Chemical Sciences, University of Canterbury, Christchurch, New Zealand}

\author{Piotr Kolenderski}
\affiliation{Institute of Physics, Faculty of Physics, Astronomy and Informatics, Nicolaus Copernicus University in Toruń, ul. Grudziądzka 5, 87-100 Toruń, Poland}

\begin{abstract}

Entanglement generated by Spontaneous Parametric Down Conversion (SPDC) involves multiple, often mutually correlated degrees of freedom.
These degrees of freedom are often treated independently, overlooking the intrinsic correlation between them. We focus on the spatial-spectral correlations that, if left uncontrolled, introduce distinguishability and reduce coherence, undermining applications such as high-dimensional OAM encoding.
We analyze the spatio-spectral structure of the biphoton and identify conditions enabling a strong reduction of such correlations.
We then quantify how spatial-spectral coupling degrades OAM spatial purity, mapping high-purity regions as functions of OAM order, crystal length, and pump/collection waists. The resulting design parameter regime enables engineering bright, high-purity OAM-entangled sources, reducing the need for additional loss-introducing filtering and therefore supporting scalable high-dimensional photonic quantum technologies.


\end{abstract}

\maketitle


\section{\label{sec:level1} Introduction} 
%







The nonlinear process of Spontaneous Parametric Down Conversion (SPDC) naturally enables the generation of entanglement across multiple degrees of freedom, including orbital angular momentum (OAM), polarization, frequency and spatial or transverse-momentum modes. 
Such a versatility carries the fundamental constraint: the energy- and momentum-conservation conditions of SPDC usually tie a photon pair’s frequencies to their emission directions \cite{grice, valencia2007, gatti, gatti1, gatti2}.
In the hyperentanglement context, more than one degree of freedom is employed and taken as independent resources \cite{hyper}, but in practice they can be correlated, so that the operations on the modal structure in a specific domain affects the accessible states in another \cite{gatti}. 
 Consequently, spatial-spectral correlations reduce the purity of the quantum states, introducing distinguishability and reducing coherence \cite{sergienko2003quantum}. 
 
Existing approaches to suppress unwanted correlations typically rely on strong filtering \cite{valencia2008, filter}.
While it improves purity, it also reduces photon flux and limits scalability in high-dimensional quantum information processing and imaging. 
What is more, a source-level design rule is lacking for a decoupling that remains valid for arbitrary pump durations. This motivates the central goal of this work: to identify experimental conditions under which spatial and spectral degrees of freedom can be engineered more independently at the source without sacrificing brightness or resource versatility (optimizing one degree of freedom does not require compromising the other, and both resources remain simultaneously available).


This challenge is specially relevant for OAM entanglement, which provides access to a high-dimensional Hilbert space, enhanced information capacity, and robustness against noise \cite{oamcom, OAMresilience}. Encoding information across multiple spatial modes supports potential high-capacity quantum key distribution (QKD) \cite{HDQKD, oamcom}, dispersion-robust quantum optical coherence tomography (QOCT) \cite{dispcancel, OAMdisp}, and improved contrast and resolution in quantum imaging \cite{contrastOAM}. The advantage of OAM entanglement in these applications, however, relies critically on spatial purity. 

\begin{figure}[ht]
    \centering
    \includegraphics[width=\columnwidth]{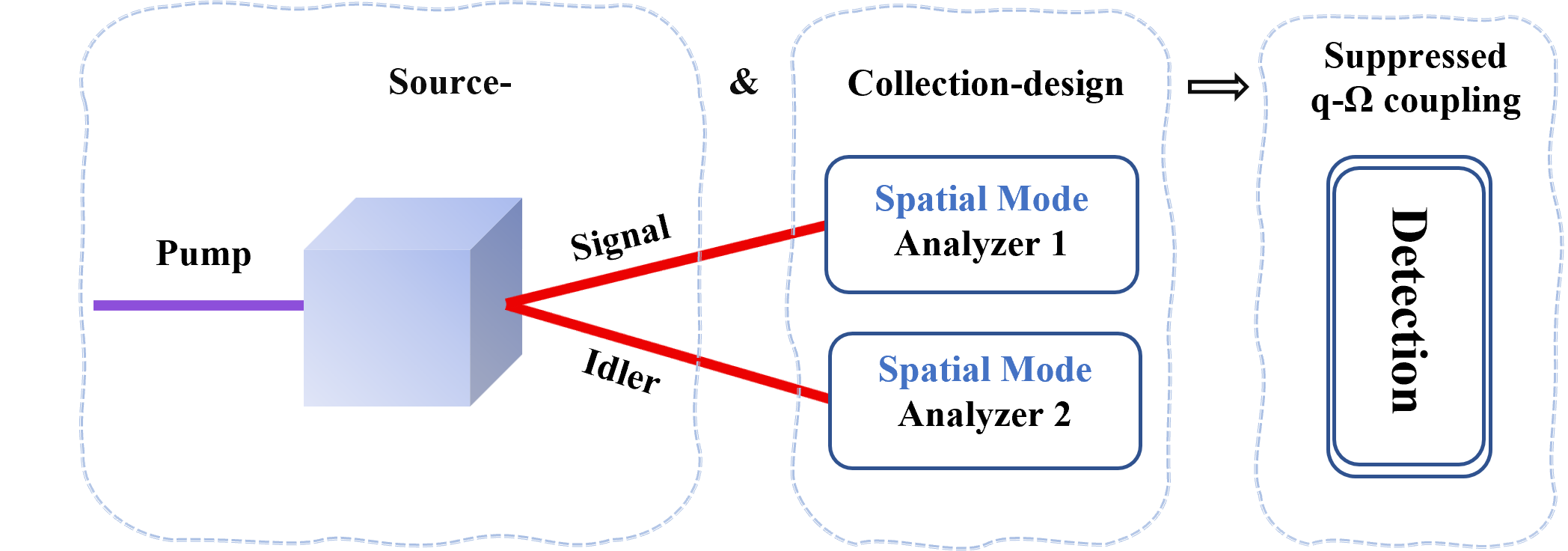}
    \caption{Conceptual schematic illustrating the combined role of source and collection parameters in selecting a SPDC regime with suppressed spatial-spectral coupling allowing high-purity OAM-entagled states. Source designing modifies the generated biphoton state through pump and phase-matching parameters, whereas in the collection stage the collection mode waist and OAM order are appropriately  chosen to optimize the purity of the detected state.}
    \label{fig:scheme}
\end{figure}

In this work, we investigate conditions that strongly suppress spatial-spectral correlations in type-I SPDC enabling high-purity OAM-entangled states generation, as illustrated in the conceptual schematic shown in \figref{fig:scheme}. We derive conditions under which the general phase-matching function factorizes into independent spatial and spectral contributions. Building on these conditions, we incorporate the pump profile to obtain a fully factorable model valid for arbitrary pulse durations.
We then compare the OAM-mode purity of the general SPDC state with that of the decoupled spatial-spectrally decoupled model, identifying regimes where the spatial purity of the general state remains high across OAM orders, crystal lengths and pump parameters.  Taken together, these results provide practical guidelines for engineering high-purity OAM photon pairs, reducing reliance on additional filtering and supporting scalable implementations in photonic quantum technologies.

The paper is organized as follows. In Sec. II we develop the spatial-spectral decoupling framework and derive the corresponding conditions, with technical details are found in Appendix~\ref{app:A} and Appendix~\ref{app:b}. In Sec. III we apply this model to OAM entanglement, analyzing the spatial purity across experimentally relevant parameters and discussing implications for source design. Finally, Sec. IV summarizes our main results and presents the conclusions.

\section{\label{sec:decouplingtypeI} SPDC Spatial-spectral decoupling}

Let us consider the type-I of a spontaneous parametric down-conversion process; under the paraxial and small-angle approximations, the dimensionless longitudinal phase mismatch can be compactly written  as \cite{gatti}: 

\begin{equation}
\begin{split}
\frac{\Delta k_z\,L}{2}
= &\biggl[
   \underbrace{\frac{|\vec{q_s}-\vec{q_i}|^2}{2\,k_p}}_{\substack{\text{transverse momentum}\\ \text{mismatch}}}
 + \underbrace{\frac{\beta}{4} \,(\Omega_s - \Omega_i)^2}_{\substack{\text{second-order}\\\text{dispersion mismatch}}}\\
 &\quad
 - \underbrace{\bigl(\tfrac{1}{v_{g,p}} - \tfrac{1}{v_{g,s}}\bigr)\,(\Omega_s + \Omega_i)}_{\substack{\text{group-velocity}\\\text{mismatch}}}
  \biggr]\frac{L}{2}.
\end{split}
\label{sincsum}
\end{equation}
Please,  refer to Appendix~\ref{app:A} for further technical details. To account for the frequency and transverse momentum dependencies of the wave vector $k=k(\Vec{q},\Omega)$, an expansion in a Taylor series is performed around small deviations from the central frequency $\Omega\ll\omega^c$ and transverse wave-vector components ($q_x$, $q_y$) around $\Vec{q}=0$.
The process occurs in a crystal of length \(L\), where the phase mismatch is given by  $\Delta k_z:=k_{z,s}+   k_{z,i}-k_{z,p}$
with $k_{z,j}$, $j\in \{p,s,i\}$, being the longitudinal component of the wavenumber of the pump, signal and idler photon, respectively. The wavevector magnitude is $k_j=\frac{n_j(\omega_j)\omega_j}{c}$, where $n_j(\omega_j)$ is the corresponding refractive index. $v_g=(\frac{\partial{k}}{\partial{\Omega}})^{-1}$ is the group velocity, where the Group Velocity Dispersion coefficient is $\beta=\frac{\partial^2{k}}{\partial^2{\Omega}}$.


Let us consider the widely used description of SPDC \cite{valencia2008,baghdasaryan-spat-temp},  
\begin{equation}
\begin{split}
        \Phi ( q_s, q_i, \Omega_s, \Omega_i)
    =\mathcal {N}&\mathrm{e}^{-\sigma_p^2|\vec{q_s}+\vec{q_i}|^2} \mathrm{e}^{-{\frac{\tau^2}{8 \ln{2}}(\Omega_s+\Omega_i)^2}}\\ & \times\sinc \biggl(  \frac{\Delta k_zL}{2}\biggr),
\end{split}
    \label{eq:phi-sinc}
  \end{equation}
where the joint amplitude is expressed as the product of the spatial and spectral Gaussian pump functions and a sinc-shaped phase-matching function, which arises from integrating the nonlinear interaction over the finite crystal length. Here, $\sigma_p=\frac{w_p}{2}$, where $w_p$ denotes the pump beam waist, and $\sigma_q$ is a spatial correlation parameter, while the pump is modeled as a pulsed Gaussian field with a finite pump pulse duration $\tau$. Although this model neglects certain technical effects (e.g., dispersion and birefringence), we shall refer to it as the general model, as it preserves the essential structure imposed by energy and momentum conservation, the key constraints on the two-photon Hilbert space.

The requirement of energy conservation during phase matching inherently couples the transverse momentum and frequency of the generated photon pair. Despite this coupling, the twin-photon state has typically been analyzed either in the spatial domain \cite{reichert2017, howell, deBritofree, debritoDS} or in the spectral domain \cite{fedorov2009, mikhailova2008, kolenderski2009}, treating these degrees of freedom separately. In this work, we consider transverse momentum and frequency generally and then we derive the conditions necessary to achieve their decorrelation.



Our approach for spatial-spectral decoupling is inspired by the work of Schneeloch and Howell  \cite{howell}, where the biphoton wavefunction was shown to be separable in the transverse momentum variables.
Under the paraxial approximation, operating in the small-argument regime, between the pump photon and the signal or idler photon allows us to further approximate the phase-matching term as $\sinc(x+ y)\approx \sinc(x) \sinc (y)$, which we then use to separate the spatial and spectral degrees of freedom.
\begin{equation}
    \sinc \left [\frac{\Delta k_z L}{2}\right]\approx\Theta(q_s,q_i)\Tilde{\Theta}(\Omega_s,\Omega_i),
    \label{sincmult2}
\end{equation}

where:
\begin{equation}
\resizebox{\columnwidth}{!}{$
    \begin{split}
  & \Theta(q_s,q_i)=\sinc \left[\frac{L}{4k_p} |\vec{q_s}-\vec{q_i}|^2\right], \\ &\Tilde{\Theta}(\Omega_s,\Omega_i)=\sinc \left[ \frac{L}{2}\left(\beta_{s(i)}(\Omega_s-\Omega_i)^2
        -\left(\frac{1}{v_{g, p}}-\frac{1}{v_{g, s}}\right)(\Omega_s+\Omega_i)\right)\right].
    \end{split}
$}
\label{sincmult}
\end{equation}
For further details on the derivation of the approximate phase mismatch refer Appendix~\ref{app:A}.

\begin{figure}[htbp!]
\centering
\subfloat[(a)]{\includegraphics[width=123pt]{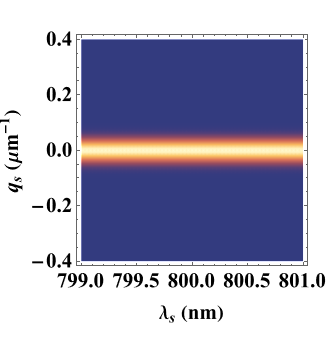}\label{fig:subfigaaa}}
\subfloat[(d)]{\includegraphics[width=123pt]{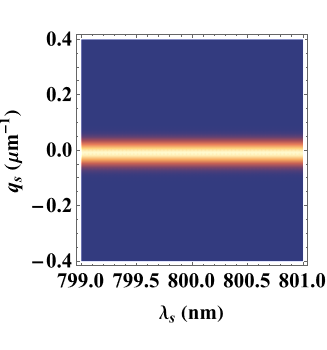}\label{fig:subfigdd}}\\[-15pt]
\subfloat[(b)]{\includegraphics[width=123pt]{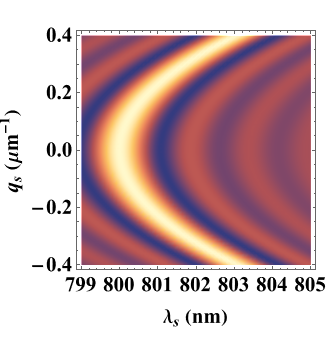}\label{fig:subfigbb}}
\subfloat[(e)]{\includegraphics[width=123pt]{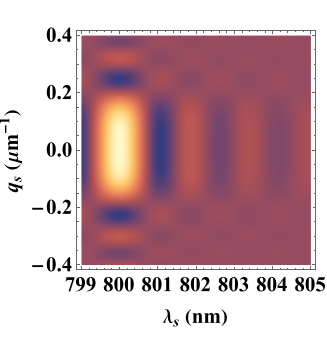}\label{fig:subfigee}}\\[-15pt]
\subfloat[(c)]{\includegraphics[width=123pt]{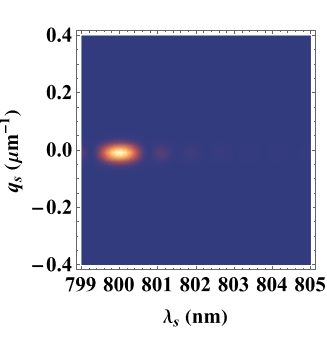}\label{fig:subfigcc}}
\subfloat[(f)]{\includegraphics[width=123pt]{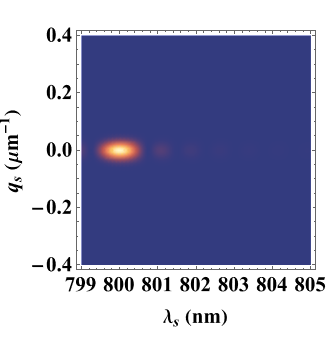}\label{fig:subfigff}}
\caption[]{Comparison of the spatial-spectral distributions predicted by the general and approximate models of Type-I SPDC. The upper panels (a–c) correspond to the general model, while the lower panels (d–f) show the approximate model. In both approaches, the same Gaussian spatial-spectral pump function is used \hyperref[fig:subfigaaa]{(a, d)}, whereas the phase-matching function is described by the full sinc expression in \hyperref[fig:subfigbb]{(b)}, given by Eq. \eqref{sincsum}, and by the double-sinc approximation in \hyperref[fig:subfigee]{(e)}, given by Eq. \eqref{sincmult2}. Panels \hyperref[fig:subfigcc]{(c, f)} show the resulting biphoton intensity distributions.
The calculations consider a pump photon with central wavelength $\lambda_p = 400\,\mathrm{nm}$, beam width $w_p = 28\,\mathrm{\mu m}$, and pulse duration $\tau=50\,\mathrm{fs}$, propagating through a Lithium Iodate (LiIO$_3$) crystal of length 
$L = 5\,\mathrm{mm}$~\cite{mikhailova2008, valencia2007}, at the phase-matching angle $\theta \simeq 42.5^\circ$.
All plots are shown as a function of the signal wavelength $\lambda_s$ and transverse momentum $q_s$ (restricted to the $x$-direction), with fixed idler wavelength $\lambda_i = 800~\mathrm{nm}$ and transverse momentum $q_i = 0.01~\mu\mathrm{m}^{-1}$.}
    \label{comparingsinc}
\end{figure}

Figure \ref{comparingsinc} compares the joint spatial-spectral structure obtained from the general and approximate type-I SPDC models. Panels \hyperref[fig:subfigaaa]{(a)}-\hyperref[fig:subfigcc]{(c)} show the spatial-spectral Gaussian pump envelope, the sinc phase-matching function (PMF) computed from \eqref{sincsum}, and the resulting joint-intensity distribution $|\Phi|^2$, respectively. 
Panels \hyperref[fig:subfigdd]{(d)}-\hyperref[fig:subfigff]{(f)} show the corresponding approximate case, where the PMF is replaced by the double-sinc approximation ~\eqref{sincmult2}-(\ref{sincmult}).
The pump function \hyperref[fig:subfigaaa]{(a, d)} defines a narrow horizontal band centered at the signal wavelength of $800~\mathrm{nm}$ for $\tau=50\,\mathrm{fs}$, extended along the spatial axis. The phase-matching function and its approximation imposes its own constraints: in the general PMF case \hyperref[fig:subfigbb]{(b)}, V-shaped structures appear (see Appendix \ref{app:c} for further details on the phase-matching geometry), while in the approximation \hyperref[fig:subfigee]{(e)} the structure is simplified to a rectangular-like modulation.

Combining the pump and phase-matching functions gives the resulting joint intensity distributions in \figref{comparingsinc}\hyperref[fig:subfigcc]{(c, f)}, showing that the approximate model captures the essential spectral selection near $\lambda_s \approx 800~\mathrm{nm}$, with good agreement for a spectral range of $\sim 1~\mathrm{nm}$. When the PMF alone is considered, it does not enforce a strong selection in transverse momentum, it mainly restricts the signal wavelength to match the idler’s, while being less discriminative in the spatial domain.

Therefore, within this range of experimental parameters a natural filtering occurs where the spatial and spectral degrees of freedom remain effectively decoupled. In other words, the spectral confinement is primarily dictated by the pump and phase-matching conditions, while the transverse momentum remains weakly constrained. This indicates that, under these approximations, correlations between frequency and transverse momentum are strongly suppressed, allowing one to treat the joint spectrum and joint spatial distributions as approximately separable contributions to the biphoton wavefunction. 

A comparison between the approximation proposed in \eqref{sincmult2}-(\ref{sincmult}) with the sinc PMF obtained from the phase mismatch in \eqref{sincsum}, considering each degree of freedom separately, is presented in  \figref{sincplots} of Appendix~\ref{app:A}. In Appendix~\ref{app:c}, we perform an analysis for a  beta-barium borate (BBO) crystal similar to the one presented here for lithium iodate (LiIO$_3$).


\captionsetup[subfloat]{justification=raggedright,singlelinecheck=false,position=top, labelformat=empty}

\subsubsection{\label{sec:dGtypeI}Approximation with Gaussians}

Building on our decomposition of the phase‐matching function into independent spatial and spectral factors, we now include the Gaussian pump envelope to write the complete two‐photon amplitude as

\begin{equation}
    \begin{split}
         \Phi ( q_s, q_i, \Omega_s, \Omega_i)=\mathcal{N} & \mathrm{e}^{-\sigma_p^2|\vec{q_s}+\vec{q_i}|^2} \mathrm{e}^{-{\frac{\tau^2}{8 \ln{2}}(\Omega_s+\Omega_i)^2}}\\&\times\Theta(q_s, q_i)  \Tilde{\Theta}(\Omega_s,\Omega_i),  
    \end{split}
\label{wft_start}
\end{equation}
where $\mathcal{N}$ is a normalization constant, $\Tilde{\Theta}$ and $\Theta$ are the spatial and spectral sinc‐multiplied phase‐matching terms in \eqref{sincmult}.
For a pump that is Gaussian both in transverse momentum and in frequency, the biphoton amplitude factorizes into a product of Gaussian envelopes, multiplied by our separable mismatch terms. By replacing each of those sinc‐shaped factors with a matching Gaussian, one arrives at a purely Gaussian wavefunction for the joint wavefunction. 
This procedure yields the fully factorable four-Gaussian approximation, whose closed-form expressions in both momentum (or position) and frequency domains will be developed next (refer to Appendix~\ref{app:A}).

 Within the simplifications considered, the spatial statistics are identical in both transverse dimensions, we restrict the analysis to the $x$-component of the transverse momenta. 
The spatial component of the phase-matching function may be approximated by a Gaussian \cite{howell}
\begin{equation}
    \Theta(q_s,q_i)\approx \mathrm{exp}\bigg[-A\bigg(\frac{L}{4k_p}\bigg)(q_{x,s}-q_{x,i})^2\bigg], 
\end{equation}
where the argument of the Gaussian matches that of the sinc function up to a constant factor $A$. The value of $A$ is chosen such that the second moments of both functions are matched.


For the spectral component $\Tilde{\Theta}(\Omega_s,\Omega_i)$,  it is essential to retain both the quadratic dispersion term and the linear group-velocity-mismatch term appearing in ~\eqref{sincmult}, as each dominates in different pulse-duration regimes. 
Whereas the spatial sinc profile admits a more straightforward Gaussian substitution, the two-argument spectral sinc requires a more involved treatment. Rather than approximating the joint wavefunction directly, one must match the reduced single-photon density matrix of the model to that of the original wavefunction by adjusting both the pump envelope and the phase-matching. 
To obtain a spectral double-Gaussian wavefunction for arbitrary pump pulse duration that approximately captures the equivalent bandwidths, we adopt (see Appendix~\ref{app:b} for details) the generalized formalism introduced in Ref.~\cite{fedorov2009}. 



The biphoton wavefunction in \eqref{wft_start} can be approximated by a separable product of Gaussian funcions in the transverse-spatial and spectral degrees of freedom:

\begin{equation}
\begin{aligned}
\Phi_\text{4G}(\vec{q_{s}},\vec{q_{i}}, \Omega_s, \Omega_i)
=\mathcal{N}_\text{4G}\,&\mathrm{e}^{-\sigma_p^2|\vec{q_{s}}+ \vec{q_{i}}|^2}\,
      \mathrm{e}^{-\sigma_q^2|\vec{q_{s}}-\vec{q_{i}}|^2} \\[6pt]
&\times \mathrm{e}^{-\alpha\,\tfrac{|\Omega_s + \Omega_i|^2}{2\,a^2(\tau)}}\,
          \mathrm{e}^{-\tfrac{(\Omega_s - \Omega_i)^2}{2\,b^2(\tau)}},
\end{aligned}
\label{4g}
\end{equation}
where $\mathcal{N}_\text{4G}$ is a normalization constant, with $\sigma_p=\frac{w_p}{2}$, where $w_p$ the pump beam waist, and \(\sigma_q\) is a spatial correlation parameter. The coefficient $\alpha$ is a dimensionless parameters introduced to improve the Gaussian representation of the general sinc-based SPDC wavefunction. Its values are set so that the squared modulus of the Gaussian approximation reproduces the central peak, while also accounting for the spectral distribution associated with the initial side lobes of the sinc-based wavefunction, which becomes more relevant in the short-pulse regime, when broader pump bandwidth allows a larger phase-matched  spectral region.
  The remaining parameters, \(a(\tau)\) and \(b(\tau)\), are defined in Appendix~\ref{app:b}, which also contains all the details for the Gaussian approximation we presented above. It also provides a  graphical comparison of \(\lvert\Phi\rvert^2\) (see Fig.~\ref{dgplots}) using the spatially and spectrally coupled wavefunction given by  ~\eqref{eq:phi-sinc}-({\ref{sincsum}}) and the the Gaussian approximation we propose in \eqref{4g}.


By adopting a factorized Gaussian form together with an optimized choice of the parameter $\alpha$ and appropriate experimental settings, our framework guarantees that both the accepted spectral bandwidth and the spatial correlations are consistent with the original spatial-spectral structure of the pump and phase-matching function. In particular, this approximation reproduces the effective bandwidth selection dictated by the SPDC geometry, without recourse to any external or artificial filtering, while preserving an approximate decoupling of the spatial and spectral degrees of freedom.

\section{\label{sec:level3} OAM-resolved Detection}

In this section, we compare the spatial purity obtained from two descriptions of the SPDC state: the spatial-spectrally coupled model (\eqref{eq:phi-sinc}-({\ref{sincsum}})) and the approximate model (\eqref{4g}), in which the spatial and spectral degrees of freedom are independent. This comparison allows us to quantify how spatial-spectral correlations affect the purity of the OAM modes collected in Laguerre-Gaussian (LG) modes.

When the phase matching conditions are fulfilled, OAM is conserved; the total OAM of the photon pair equals that of the pump, so the down-converted photons emerge entangled in this degree of freedom. Thus, for the Gaussian pump considered here, $\ell_p=0$, the signal and idler axial OAM number obey the relation $\ell_s=-\ell_i=\ell$. 
In this section, we investigate the spatial purity of the spatial-spectral correlated SPDC state, where we model the spatial collection modes of the signal and idler photons as Laguerre-Gaussian modes, which is among the most widely studied OAM-carrying beams.

 Following the approach presented in Ref. \cite{valencia2008}, in which spatial filtering is introduced into the spatiotemporal mode function through multiplicative spatial collection functions, we consider the spatial-spectral mode function 
\begin{equation}
\begin{split}
\Phi_{LG}(\vec{q}_s,\vec{q}_i,\Omega_s, \Omega_i)
= 
  \Phi(\vec{q}_s,\vec{q}_i,\Omega_s, \Omega_i)\,
  &LG_{p_s}^{\ell_s}(\vec{q}_s) LG_{p_i}^{\ell_i}(\vec{q_i}).
\end{split}
\end{equation}
 Here, $\Phi$ denotes the wavefunction of the down-converted photons, while the functions $LG_{p_s}^{\ell_s}$ and $LG_{p_i}^{\ell_i}$ describe the Laguerre-Gaussian collection modes of the signal and idler photons, respectively. The generic LG mode is given by
\begin{equation}
\begin{split}
  LG_p^{\ell}(q, \phi)=&\sqrt{\frac{p! \, w_0^2}{4\pi (|\ell|+p)!}}\;
2^{\tfrac{|\ell|}{2}+\tfrac{1}{2}}
\left(\frac{q w_0}{2}\right)^{|\ell|}
L_p^{|\ell|}\!\left(\frac{w_0^2 q^2}{2}\right)\\ &
\exp\!\left(-\frac{w_0^2 q^2}{4}\right)
e^{i\ell\phi},
  \end{split}
\end{equation}
expressed in cylindrical coordinates $\vec{q}=$($q$, $\phi$).
Here $w_0$ is spatial collection mode waist, and $L_p^{|\ell| }(y)$ is called the associated Laguerre polynomials.

The analysis presented in this section is based on two different forms of $\Phi$, corresponding to two different regimes: (1) when the spatial and spectral degrees of freedom are coupled, we use \eqref{eq:phi-sinc}-({\ref{sincsum}}); and (2) when these degrees of freedom are independent, we use \eqref{4g}.

    
Considering the  correlations, we regard the spatial and spectral degrees of freedom as two subsystems of the biphoton state. To quantify the spatial purity, we trace out the spectral subspace from the  density operator of the full system \( \rho = \ket{\Phi}_{s,i} \bra{\Phi}_{s,i} \). This yields $
\hat \rho_q=  \int \mathrm{d}\Omega_s'' \, \mathrm{d}\Omega_i'' \, \langle \Omega_s'', \Omega_i'' | \rho | \Omega_s'', \Omega_i'' \rangle$, which in terms of the joint spatial-spectral wavefunction, becomes \cite{valencia2008}
\begin{equation}
\begin{split}
\hat \rho_q = \int  \mathrm{d}\vec{q_s}& \mathrm{d}\Omega_s \mathrm{d}\vec{q_i} \mathrm{d}\Omega_i \mathrm{d}\vec{q_s'} \mathrm{d}\vec{q_i'}\Phi_{LG}^*(\vec{q_s}',  \vec{q_i}', \Omega_s, \Omega_i)
 \\ \times &\Phi_{LG}(\vec{q_s},  \vec{q_i}, \Omega_s, \Omega_i) 
| \vec{q}_s, \vec{q_i} \rangle \langle \vec{q_s}', \vec{q_i}' |.
\end{split}
\label{redrho}
\end{equation}

Finally, the spatial purity is defined as
\begin{equation}
P = \mathrm{Tr}(\tilde{\rho_q}^2),
\end{equation}
Since the biphoton amplitude is defined up to an overall normalization factor, the reduced density matrix is normalized to unit trace $\tilde{\rho_q}=\frac{\hat{\rho_q}}{\mathrm{Tr}(\hat{\rho_q})}$. 

\subsection{Discussion}

The spatial purity $P$ as a function of the ratio between the collected-photons beam waist $w_{s,i}$ and the pump beam waist $w_p$ is shown in Fig \ref{fig:purity1}. Note that we assume signal and idler have the same collection beam waist size $w_{s}=w_{i}$. To evaluate the purity, we simplify the phase-mismatch function \eqref{sincsum}, by retaining only the dominant quadratic terms in both spatial and spectral variables. This gives the essential spatial-spectral correlation structure while making the integrals analytically tractable. For further simplification, we restricted the analysis to the case of radial quantum number \( p = 0 \).

\def\ccc{115pt}
\begin{figure}[ht]
\centering
    \subfloat[][
   (a) $\ell=4$, $L=5\,\mathrm{m m}$
    ]{
       \includegraphics[width=\ccc]{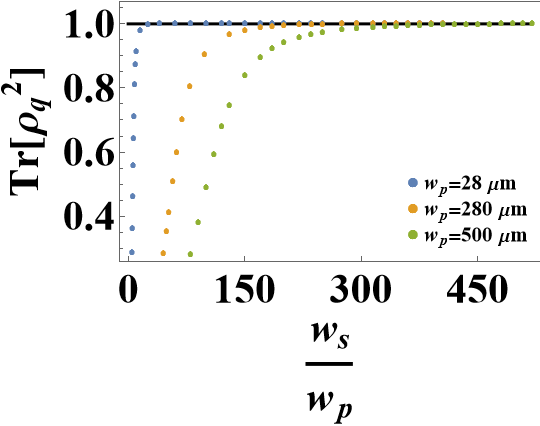} \label{fig:subfigw}
    }
         \subfloat[][ (b)
    $\sigma_p=28\,\mathrm{\mu m}$, $L=5\,\mathrm{m m}$     
    ]{
     \includegraphics[width=\ccc]{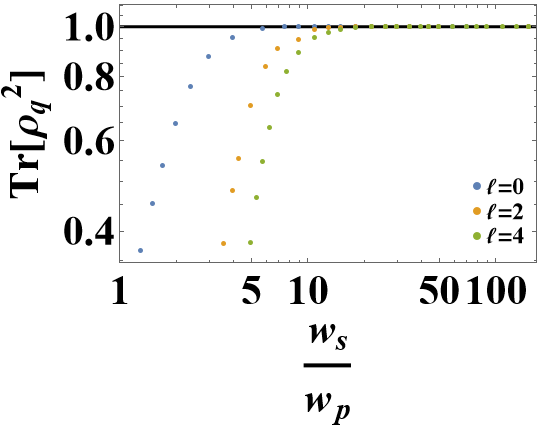} \label{fig:subfigl}
     }\\
 \subfloat[][ (c)
    $\sigma_p=28\,\mathrm{\mu m}$, $\ell=4$\newline
    ]{
        \includegraphics[width=\ccc]{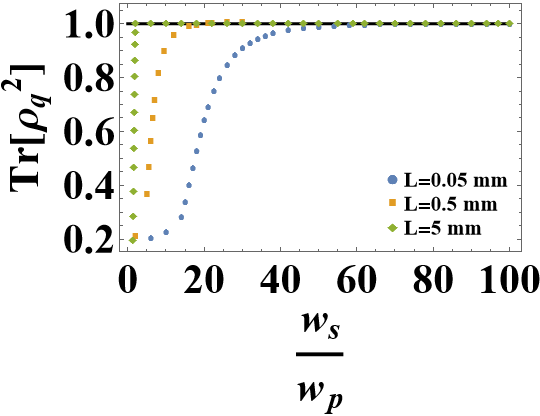} \label{fig:subfigll}
    } 
      \subfloat[][(d)
    $\sigma_p=28\,\mathrm{\mu m}$, $\ell=4$, $L=5\,\mathrm{m m}$
    ]{
     \includegraphics[width=\ccc]{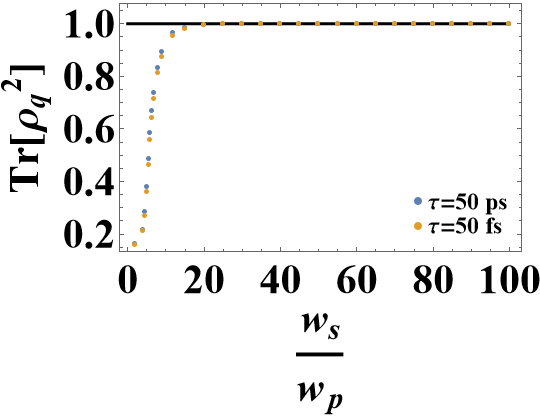} \label{fig:subfigt}
}
\caption[Optional caption for list of figures]{The spatial purity for the SPDC general model  \eqref{eq:phi-sinc} (curves with symbols) and for the four-Gaussian model \eqref{4g} (solid curve) are plotted as a function of the ratio between the spatial collection mode waist $w_{s}$ ($w_{i}=w_{s}$) and the pump beam waist $w_p$. The purity provided by the spatial-spectral coupled wavefunction is compared with the one of the separable  Gaussian model varying \subref{fig:subfigw} the OAM number $\ell_s=-\ell_i=\ell$, \subref{fig:subfigl} crystal length, \subref{fig:subfigll} pump beam waist and pump pulse duration \subref{fig:subfigt}. For each model, the reduced density matrix is independently normalized to unit trace before evaluating the purity.}
\label{fig:purity1}
\end{figure}

For the SPDC model in \eqref{eq:phi-sinc}, where the phase-matching function couples spatial and spectral degrees of freedom, the spatial purity can deviate significantly from unity. As shown in \figref{fig:purity1}\subref{fig:subfigw}, the purity drops for small values of the ratio $w_s/w_p$ and approaches unity as this ratio increases. Because a mode with waist $w$ produces an angular (transverse-momentum) distribution proportional to $1/w$, a smaller collection waist (small $w_s$) yields a broader angular acceptance, detecting photons whose transverse-momentum components are more strongly coupled to their frequency mode. Conversely, enlarging the collection waist narrows the width in transverse momentum, thereby collecting modes with suppressed spatial-spectral correlations.


The role of the axial OAM number is also evident in \figref{fig:purity1}\subref{fig:subfigl}: higher indices $\ell$ require larger $w_s$ to achieve the same purity. This arises because higher-$\ell$ LG modes extend further in momentum space, sampling larger $\vec{q}$ values where spatial-spectral correlations are stronger.
Panel \figref{fig:purity1}\subref{fig:subfigll} shows the spatial purity for different values of the crystal length $L$, note that the purity rises faster as a function of $w_s$ for longer crystal. Increasing the length of the crystal tightens the phase-matching cone, it admits a smaller spread of emission angles, therefore fewer angle-dependent frequency components entering the collected mode.

Finally, the purity is insensitive to the pump pulse duration \figref{fig:purity1}\subref{fig:subfigt} for wavelength-degenerate photons. In this case, the twin photons can populate any transverse position essentially independent of the pump pulse duration. 
Since the four-Gaussian wavefunction (solid black curve) is a separable state in the spatial and spectral degrees of freedom, tracing over the frequency leaves the spatial state pure.

\section{Conclusion}

We have investigated the conditions under which spatial-spectral correlations in SPDC can be strongly suppressed and established a factorable Gaussian model for the resulting approximately separable state. We then identify the source parameters and collection configurations for which the sinc-based SPDC state approaches this factorable regime, for both short- and long-pulse pumping, and examine the corresponding spatial purity of the collected OAM modes. 
Instead of providing a single set of parameters, our results identify experimentally relevant parameter regimes that can guide the choice of source and collection conditions to minimize spatial-spectral coupling for high-purity OAM modes.

Reliable generation of high-purity OAM-entangled states is an important step towards their broader use in quantum photonic technologies.
OAM modes are attractive for high-dimensional photonic protocols, where their advantages depend critically on high modal purity and stable mode matching.
Therefore, to avoid restrictive filtering that reduces the performance of the source, we aim to enhance the collection efficiency of high-purity OAM-entangled states generated. 




\begin{acknowledgments}
The authors acknowledge the financial support from Horizon Europe, the European Union's Framework Programme for Research and Innovation, SEQUOIA project, under Grant Agreement No. 101070062.
SMK acknowledges New Zealand Ministry of Business, Innovation and Employment (MBIE) Smart Ideas funding (E7943). 
\end{acknowledgments}

\appendix

\section{Derivation of the approximate Phase Mismatch $\Delta k_z$ }
\label{app:A}

\begin{center}\textbf{\footnotesize  Type-I SPDC }
\end{center}
\vspace{0.5cm}

We derive here the expression for the approximated longitudinal phase mismatch Eq. (\ref{sincsum}), in the context of spontaneous parametric down-conversion, incorporating both spatial and spectral contributions.


The longitudinal phase mismatch is defined as:
\begin{equation}
        \Delta k_z= k_{p,z} - k_{s,z} - k_{i,z}.
        \label{mismatchk}
\end{equation}
$k_{j,z}$ denotes the longitudinal component of the wavevector for the pump ($j=p$), signal ($j=s$), and idler ($j=i$) photons. The spatial-spectral wavevector (to second order) can be written as \cite{gatti}
\begin{equation}
    k_{j,z}\approx k_j-\frac{|\vec{q}_j|^2}{2k_j}+ \frac{\Omega_j}{v_{g,j}} + \frac{\beta }{2}\Omega_j^2,
    \label{fullveck}
\end{equation}
where the transverse walk-off is set to zero  $\alpha =\left(\frac{\partial{k}}{\partial{\vec{q}}}\right)=0$ as an additional simplifying assumption.


\begin{itemize}
\item Spatial Contribution
\end{itemize}

The momentum conservation can be broken down into its longitudinal and transverse components, as follows \cite{howell}

\begin{equation}
\begin{split}
       \Delta k_z^{(\text{spatial})} = k_p-(k_s+k_i) \cos{\theta},
\end{split}
\label{PMlong}
\end{equation}

\begin{equation}
\begin{split}
        k_s \sin\theta \approx \frac{|\vec{q}_s - \vec{q}_i|}{2}.
\end{split}
\label{PMtransv}
\end{equation}
 For signal and idler photons emitted at small angles
$\theta$ relative to the pump propagation direction, we can apply the small-angle approximation, and substituting the last equation \eqref{PMtransv} into the first one \eqref{PMlong}, we obtain \cite{howell}:
\begin{equation}
       \Delta k_z^{(\text{spatial})}\approx \frac{|\vec{q}_s - \vec{q}_i|^2}{2k_p}.
       \label{spatdeltak}
\end{equation}


\begin{itemize}
\item Spectral Contribution
\end{itemize}

The frequency dependence of the phase mismatch is \cite{fedorov2009}

\begin{align*}
\Delta k_z^{\mathrm{(spec)}} &= k_p + \frac{\Omega_s + \Omega_i}{v_{g,p}} + \frac{\mathrm{\beta}_p}{2}(\Omega_s + \Omega_i)^2 \\
&\quad - \left(k_s + \frac{\Omega_s}{v_{g,s}} + \frac{\mathrm{\beta}_s}{2}\Omega_s^2\right) \\
&\quad - \left(k_i + \frac{\Omega_i}{v_{g,i}} + \frac{\mathrm{\beta}_i}{2}\Omega_i^2\right).
\end{align*}
The superscript \textit{spec} stands for spectral.

Note that in type-I SPDC, the signal and idler photons share identical polarization and thus experience the same refractive index in the nonlinear crystal. This results in identical group velocities $(v_{g, s}=v_{g, i})$ and group velocity dispersion coefficients $(\beta_s=\beta_i)$, simplifying the combined spectral term to

\begin{equation*}
\begin{aligned}
 \Delta k_z^{(\text{spec})} &=  k_p - k_s - k_i +\left( \frac{1}{v_{g, p}} - \frac{1}{v_{g, s}} \right)(\Omega_1 + \Omega_2) \\
&\quad + \frac{1}{2} \left[ - \beta_s(\Omega_1^2 + \Omega_2^2)  +\beta_p (\Omega_1 + \Omega_2)^2\right]. 
\end{aligned}
\end{equation*}
We neglect the contribution of dispersion from the pump wavevector, the last term in the above equation, since it constitutes only a minor correction. This contribution is already captured by the linear group velocity mismatch, as it involves the same frequency combination $\Omega_s + \Omega_i$.
Under this approximation, the quadratic part can be further simplified by rewriting $\Omega_s^2 + \Omega_i^2 = \frac{1}{2}[(\Omega_s + \Omega_i)^2 + (\Omega_s - \Omega_i)^2] \to \frac{1}{2}(\Omega_s - \Omega_i)^2$.
 The spectral phase mismatch then becomes \cite{fedorov2009}

\begin{equation}
\Delta k_z^{(spec)} \approx  \frac{\beta_{s}}{\textbf{4}}(\Omega_s - \Omega_i)^2 
- \left( \frac{1}{v_{g, p}} - \frac{1}{v_{g, s}} \right)(\Omega_s + \Omega_i).
\label{deltakspec}
\end{equation}
One might ask why the quadratic term is kept, given that it is typically smaller than the linear term. It cannot be neglected as it ensures a finite-width single-photon spectrum. Conversely, the linear term can be suppressed for long pump pulses (e.g. in the continuous-wave regime), where \( \Omega_s \approx -\Omega_i \), thereby eliminating its contribution. However, the linear term is essential  for short pulses and cannot be ignored in that regime.

 
\medskip

Combining both spatial and spectral contributions yields the final approximated expression for the longitudinal phase mismatch of the SPDC type-I \cite{baghdasaryan}:
\begin{equation}
\begin{split}
\Delta k_z \approx & -\frac{|\vec{q}_s - \vec{q}_i|^2}{2k_p} 
+ \frac{\mathrm{\beta}_{s(i)}}{4}(\Omega_s - \Omega_i)^2 
\\ &- \left( \frac{1}{v_{g, p}} - \frac{1}{v_{g, s}} \right)(\Omega_s + \Omega_i).
\end{split}
\end{equation}

\noindent
It highlights the interplay of spatial and spectral correlations in SPDC under the combined small-angle and narrowband approximations. 
The derivation systematically incorporates second-order dispersion and transverse momentum conservation, providing a compact analytic form suitable for further theoretical and experimental modeling.

\def\ccc{115pt}
\begin{figure}[ht]
    \subfloat[][
    (a) $q_s=-q_i=0.01\,\mathrm{\mu m^{-1}}$ \newline
\phantom{=}\,$\lambda_i=800\,\mathrm{nm}$]
    {\hspace{-0.4cm}\includegraphics[width=4.3cm]{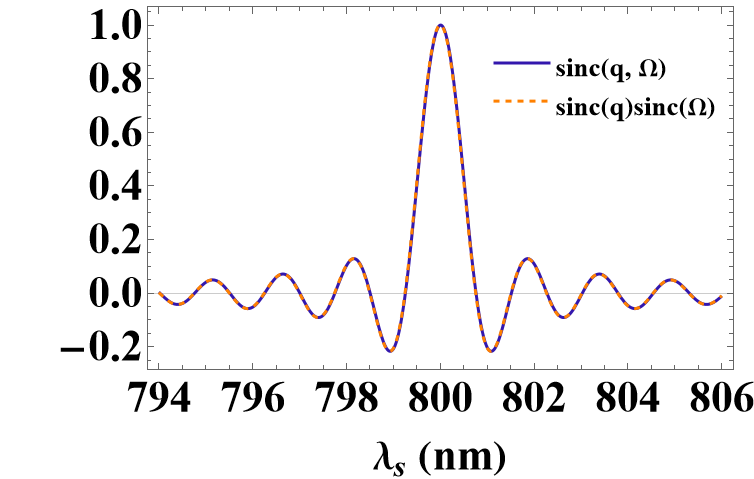}
    \label{fig:subfigaa}
    } 
    \subfloat[][
    (b) $\lambda_s=\lambda_i=800\,\mathrm{nm}$ \newline
   \phantom{====}$q_i= 0.01\,\mathrm{\mu m^{-1}}$
    ]{
    \includegraphics[width=4.3cm]{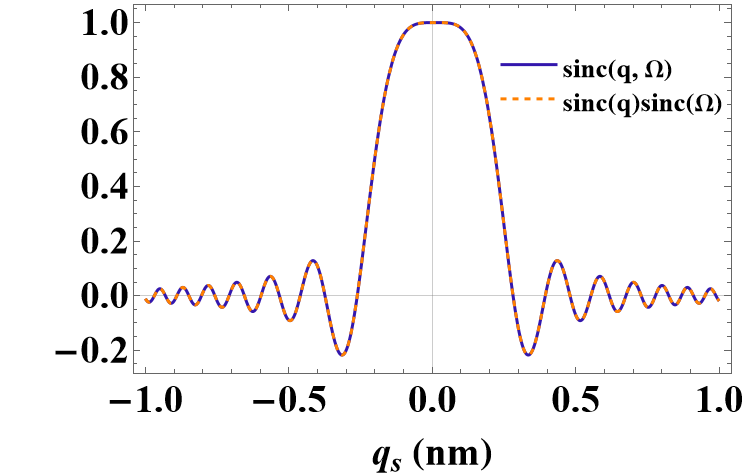}
    \label{fig:subfigdd}
    }
\caption[Optional caption for list of figures]{ Phase matching function for semi-general model obtained from \eqref{sincsum} (solid blue line) and its double-sinc approximation represented by \eqref{sincmult2}-(\ref{sincmult}) (dashed orange line). The double-sinc function is a good approximation of the phase matching function for a broad range of wavelengths \subref{fig:subfigaa}, if the relative transverse momentum   between the photons is small $|q_s-q_i| \approx 0.02 \,\mathrm{\mu m^{-1}}$. The double-sinc approximation also fits well the behavior of the general PMF in the spatial domain; for wavelength degenerated photons \subref{fig:subfigdd}, the approximation covers the same wide range of transverse momenta allowed by the general phase-matching conditions. Results obtained for a SPDC type-I photons exiting a LiIO3 crystal with a length of $L=5\,\mathrm{mm}$, pump wavelength $\lambda_p=400\,\mathrm{nm}$ and beam waist $\sigma_p=28\,\mathrm{\mu m}$. }
\label{sincplots}
\end{figure}

We analyze how the approximation we propose in \eqref{sincmult2}-(\ref{sincmult}) compares to the general sinc phase matching function obtained from \eqref{sincsum} in \figref{sincplots}. Assuming an idler photon wavelength of 800$\,\mathrm{nm}$, the double-sinc approximation (dashed orange curve) fits well the general phase matching function (solid blue curve) for a relatively large range of signal photon wavelengths  \subref{fig:subfigaa}, as long as the difference between signal and idler transverse momentum is not much bigger than $|q_s-q_i| \approx 0.02 \,\mathrm{\mu m^{-1}}$.  \subref{fig:subfigdd} Similarly, the curves align nicely for a wide range of transverse momentum of the signal photon $q_s$  under the condition that the twin photons are wavelength degenerated. In summary, when the difference between the transverse momentum of the photons is small, there is a relatively loose restriction in the wavelength range allowed for the photons, and as long as the photons are of similar wavelength, approximately no limitation is applied to their transverse momentum. 
This reflects the PMF conditions under which the spatial and spectral degrees of freedom are decoupled. In Appendix~\ref{app:c}, we perform an analysis for a BBO crystal similar to the one presented here for lithium iodate (LiIO$_3$).

\begin{center}\textbf{ \footnotesize Type-II SPDC }
\end{center}
\vspace{0.3cm}

The SPDC type-II is characterized by associating orthogonal polarization to the signal and idler beams, with one of them having the same polarization as the pump beam. Consider that the pump beam propagates as an ordinary wave, and the signal and idler beams as ordinary and extraordinary waves ($o\rightarrow o+e$), respectively.

The wavevector mismatch for SPDC type-II is written as
\begin{equation}
\begin{split}
        \Delta k_z=
        &\bigg(\frac{1}{v_g^p}(\Omega_s+\Omega_i)-\frac{1}{v_g^{s}}\Omega_s-\frac{1}{v_g^{i}}\Omega_i\bigg)\\
        &+\frac{\beta_{p}}{2}(\Omega_s+\Omega_i)^2 -\frac{\beta_{s}}{2}\Omega_s^2-\frac{\beta_{i}}{2}\Omega_i^2\\
        &-\frac{|\vec{q}_s|^2}{2k_s}-\frac{|\vec{q}_i|^2}{2k_i},
        \end{split}
        \label{coupleII}
\end{equation}
which is obtained from the longitudinal wavevector expansion \eqref{fullveck}, substituting the pump, signal and idler contributions into the phase mismatch \eqref{mismatchk}, using energy conservation and assuming perfect phase matching at the central frequencies.

Assuming that the idler photon propagates along the principal axis of the crystal, we discard, without loss of generality, the spatial walk-off of the extraordinary beam ($\alpha =(\frac{\partial{k}}{\partial{\vec{q}}})=0$).
Therefore, the spatial contribution for the phase mismatch for type-II is the same as for type-I \eqref{spatdeltak}.

\begin{equation}
       \Delta k_z^{(\text{spatial})}\approx \frac{|\vec{q}_s - \vec{q}_i|^2}{2k_p}.\nonumber
\end{equation}

For the spectral contribution, we adopt here a strategy similar to that used for type-I SPDC, where we have dropped the pump temporal dispersion term \(\frac{\beta_p}{2}(\Omega_s+\Omega_i)^2\), keeping the linear term with the same dependence.
The spectral counterpart of the mismatch for type-II reads as 
\begin{equation}
\begin{split}
  \Delta k_z^{spec}\approx&   -\frac{(\sqrt{\beta_{s}}\Omega_s-\sqrt{\beta_{i}}\Omega_i)^2}{2}+\\ 
    &+\frac{1}{v_g^p}(\Omega_s+\Omega_i)-\frac{1}{v_g^{s}}\Omega_s-\frac{1}{v_g^{i}}\Omega_i\Bigg),
    \end{split}
\end{equation}
where we have used the approximation 

\begin{equation}
    \beta_s\Omega_s^2 + \beta_i\Omega_i^2
      \approx \tfrac12(\sqrt{\beta_s}\,\Omega_s-\sqrt{\beta_i}\,\Omega_i)^2,
\end{equation}
where we neglected the first term on the right side of $  \beta_s\Omega_s^2 + \beta_i\Omega_i^2:=\tfrac12\big[(\sqrt{\beta_s}\,\Omega_s+\sqrt{\beta_i}\,\Omega_i)^2+(\sqrt{\beta_s}\,\Omega_s-\sqrt{\beta_i}\,\Omega_i)^2\big]$, valid in the region where $\beta_s\,\Omega_s\approx-\beta_i\,\Omega_i$.

Combining both spatial and spectral contributions yields the final approximated expression for the longitudinal phase mismatch of the SPDC type-II:
\begin{equation}
\begin{split}
\Delta k_z \approx & -\frac{|\vec{q}_s - \vec{q}_i|^2}{2k_p} 
+ \frac{(\sqrt{\beta_{s}}\Omega_s-\sqrt{\beta_{i}}\Omega_i)^2}{4}+\\ 
    &-\frac{1}{v_g^p}(\Omega_s+\Omega_i)+\frac{1}{v_g^{s}}\Omega_s+\frac{1}{v_g^{i}}\Omega_i.
\end{split}
        \label{sincsII}
\end{equation}



\section{Biphoton's Gaussian Modelling}
\label{app:b}

\begin{center}\textbf{ \footnotesize Type-I SPDC }
\end{center}
\vspace{0.3cm}

We analyze in more detail how the spatial and spectral degrees of freedom contributes to the structure of the biphoton state, using the separable approximation in \eqref{sincmult} for the sinc phase-matching function obtained from \eqref{sincsum}.

Let us start by writing the two-photon wavefunction for a spectral and spatial Gaussian pump
\begin{equation}
\begin{split}
    \Phi ( \vec{q_s}, \vec{q_i}, \Omega_s, \Omega_i)&=  \Phi(\Omega_s,\Omega_i)\Phi( \vec{q_s}, \vec{q_i}),
\end{split}
\label{tI}
\end{equation}    
where
\begin{equation}
\begin{split}
\Phi( \vec{q_s}, \vec{q_i})=&\mathrm{e}^{-w_p^2|\vec{q_s}+\vec{q_i}|^2}  \sinc \left[\frac{L}{4k_p} |\vec{q_s}-\vec{q_i}|^2\right]
\end{split}
\label{tIq}
\end{equation}    
and
\begin{equation}
\begin{split}
&\Phi(\Omega_s,\Omega_i)
=\exp\!\Bigl[-\tfrac{\tau^2}{8\ln2}\,(\Omega_s+\Omega_i)^2\Bigr]\;\\ &\times
\sinc\!\Bigl[\tfrac{L}{2}\Bigl(\tfrac{\beta_s}{4}\,(\Omega_s-\Omega_i)^2
-\bigl(\tfrac{1}{v_{g,p}}-\tfrac{1}{v_{g,s}}\bigr)\,(\Omega_s+\Omega_i)\Bigr)\Bigr].
\label{phispec}
\end{split}
\end{equation}

The spatial phase matching function may be approximated by a Gaussian \cite{howell}:
\begin{equation}
    \sinc\bigg[\frac{L}{4k_p} (q_{x_s}-q_{x_i})^2\bigg] \approx \mathrm{exp}\bigg[-A\bigg(\frac{L}{4k_p}\bigg)(q_{x_s}-q_{x_i})^2\bigg],\nonumber
\end{equation}
Assuming that the argument of the sinc function matches that of the Gaussian, except for a constant prefactor $A=\frac{8}{3}$, which ensures that both functions yield the same second moments when integrated over the reduced probability density. 




 To model the spectral structure across arbitrary temporal pump profiles, we follow the framework developed by Fedorov and collaborators in Ref.~\cite{fedorov2009}.
For long pulses  (\( \Omega_s \approx -\Omega_i \)), the sinc function can be well approximated by a Gaussian, and the joint spectral amplitude \( \Phi(\Omega_s, \Omega_i) \) admits a direct double-Gaussian form. However, in the short-pulse regime,  the coexistence of both linear and quadratic spectral terms in  \eqref{phispec} leads to an asymmetric sinc function for which a direct Gaussian approximation is inaccurate. Instead, a Gaussian is fitted to the reduced spectral density instead $\rho_s(\Omega_s, \Omega_s') = \int d\Omega_i\, \Phi(\Omega_s, \Omega_i)\Phi^*(\Omega_s', \Omega_i)$, and by treating the asymptotic regimes and interpolating between them, a general double-Gaussian spectral wavefunction describes well the state for any pump pulse duration.

Under the approximations listed above, the spatially-spectrally separable wavefunction for type-I SPDC takes the four-Gaussian form 
\begin{equation}
\begin{split}
   \Phi ( x_s, x_i, \Omega_s, \Omega_i)=\mathcal{N} & \, \mathrm{e}^{-\frac{(x_s+x_i)^2}{16\sigma_p^2}} \, \mathrm{e}^{-\frac{(x_s-x_i)^2}{16\sigma_x^2}} \\
   & \times \mathrm{e}^{-\alpha\frac{(\Omega_s+\Omega_i)^2}{2a^2(\tau)}} \, \mathrm{e}^{-\frac{(\Omega_s-\Omega_i)^2}{2b^2(\tau)}},\nonumber
\end{split}
\end{equation}
expressed in the transverse-position representation, obtained by inverse Fourier transforming the transverse-momentum counterpart in \eqref{4g}.
$\mathcal{N}$ is a normalization constant, $\sigma_p=\frac{w_p}{2}$,  $w_p$ is the pump beam waist. The position correlation width parameter is defined as \( \frac{1}{16\sigma_x^2} = \frac{3k_p}{8L} \), where $k_p=\frac{2 \pi}{\lambda_p} n_{e,eff}(\lambda_p,\theta)$, $\theta$ is the angle between the pump and the crystal axis. We set \( \alpha = 0.4 \) for short pump pulses and \( \alpha = 1 \) for long pump pulses, see \figref{dgplots}(b) and (c).



\noindent
 The effective spectral widths are given by \cite{fedorov2009}
\begin{equation}
\begin{split}
a(\tau)& = \frac{1.39}{\pi A \sqrt{\ln 2}} \frac{\lambda_0}{L} (1 + \eta^\gamma)^{-1/\gamma} \omega_0, \\ \quad
b(\tau)& = \sqrt{\frac{\lambda_0}{2\pi \cdot 0.249 \cdot B L}} \cdot \frac{(1 + \eta^\gamma)^{1/2\gamma} \omega_0}{\sqrt{\eta}},
\end{split}
\label{ab}
\end{equation}
where the auxiliary parameters are defined as follows:
\begin{equation}
\begin{split}
\text{A}& =\left( \frac{1}{v_{g, p}} - \frac{1}{v_{g, s}} \right)^{-1}, \quad 
\eta= \frac{2c\tau}{AL}, \quad \\ 
\text{B}& = \frac{\omega_0 c \, \beta_{s}}{4}, \quad 
\gamma = 2.21.
\end{split}
 \label{AB}
\end{equation}
Recall that \( \omega_0 \) denote the central frequency of the pump spectrum,  $c$ is the speed of light, \( \tau \) is the pump pulse duration and \( L \) is the crystal length.

To investigate whether our four-Gaussian model, while analytically simple but not simplistic, accurately captures the decoupled spatial-spectral structure of the general biphoton wavefunction, we compare it with the general model under representative experimental conditions.
Figure~\ref{dgplots} illustrates the squared modulus of the type-I SPDC  general wavefunction (solid blue curve), given by a spatial-spectral Gaussian pump functions and the PMF \eqref{eq:phi-sinc}-(\ref{sincsum}), and the approximation (dashed orange curve) given by the four-Gaussian introduced in \eqref{4g}. 

Panel \figref{dgplots}\subref{fig:subfigag} shows the spatial profile of the biphoton amplitude as a function of the signal photon’s transverse position $x_s$, with the idler fixed at $x_i = 1\,\mu\mathrm{m}$ and both photons at the wavelength $\lambda_s = \lambda_i = 800\,\mathrm{nm}$. Notice that, within the conditions we set, the  probability density obtained  from the four-Gaussian  wavefunction fairly agrees with the one from the general wavefunction over a wide spatial range. We observed the same spatial profiles for $\tau = 50\,\mathrm{ps}$ and $\tau = 50\,\mathrm{fs}$, that is why we show only one case here. This result is consistent with the fact that the pump pulse duration should not affect the spatial component of the state in a regime where the two degrees of freedom are decoupled.

A long pump pulse yields a narrow pump spectral bandwidth, whereas a short pump pulse has a broader spectrum and therefore permits a wider range of signal and idler wavelength combinations. The finite pump bandwidth is explicitly included in the sinc-based wavefunction \eqref{eq:phi-sinc}-(\ref{sincsum}) through the Gaussian spectral envelope determined by the pump-pulse duration $\tau$. In the approximate wavefunction \eqref{4g}, this dependence is incorporated into the parameter a($\tau$).
\subref{fig:subfigbg} In the spectral domain, we note a relatively good agreement between the curves if the photons are perfectly indistinguishable in wavelength for a long pump pulse duration ($\tau=50\,\mathrm{ps}$), and photon's separation of $2\,\mathrm{\mu m}$.  \subref{fig:subfigcg} For a short pump pulse duration ($\tau=50\,\mathrm{fs}$), the spatial and spectral degrees of freedom remain decoupled even for a certain bandwidth, our model covers this behavior by adjusting the parameter $\alpha$ in Eq. (\ref{4g}). In Appendix~\ref{app:c}, we perform an analysis for a BBO crystal similar to the one presented here for LiIO$_3$.


\def\ccc{115pt}
\begin{figure}[ht]
\centering
    \subfloat[][
    (a) $\lambda_s=\lambda_i=800\,\mathrm{nm}$ \newline
     \phantom{=}\,$x_i=1\,\mathrm{\mu m}$
    ]{
    \includegraphics[width=130pt]{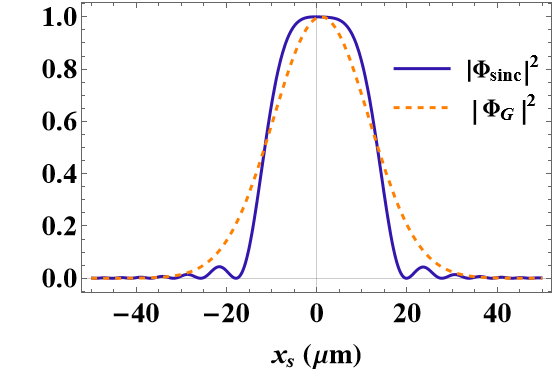}
    \label{fig:subfigag}
    }    \newline
    \subfloat[]
    [
     (b) $\tau=50\,\mathrm{ps}$, $\lambda_i=800\,\mathrm{nm}$\newline
     \phantom{=}\,\phantom{=}\,$x_s=-x_i=1\,\mathrm{\mu m}$ 
    ]
    {\includegraphics[width=\ccc]{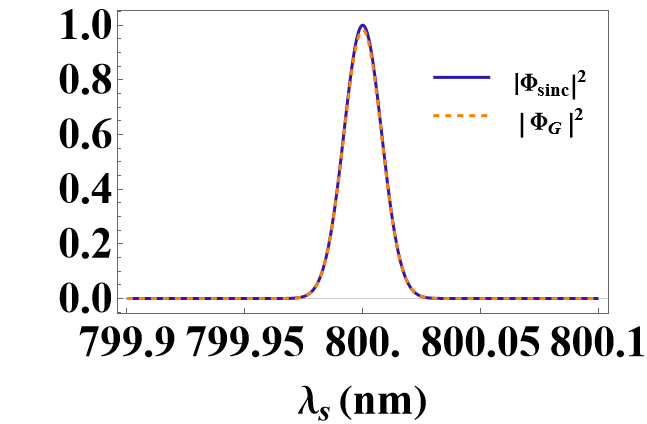}
    \label{fig:subfigbg}
    }
    \subfloat[][
    (c) $\tau=50\,\mathrm{fs}$,  $\lambda_i=800\,\mathrm{nm}$ \newline
     \phantom{=}\,\phantom{=}\,$x_s=-x_i=1\,\mathrm{\mu m}$ 
    ]{
    \includegraphics[width=\ccc]{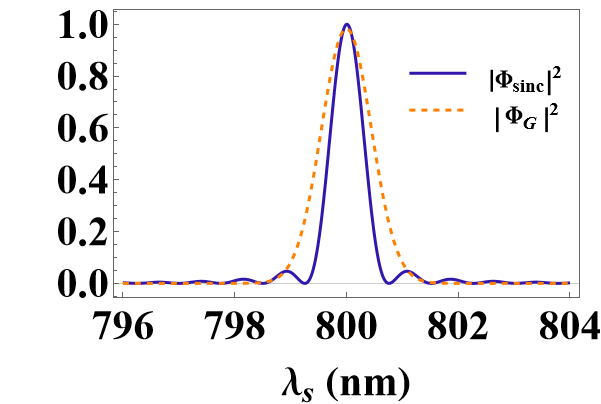}
    \label{fig:subfigcg}
    }

\caption[Optional caption for list of figures]{ The squared modulus of the SPDC type-I  general wavefunction \eqref{eq:phi-sinc}-(\ref{sincsum}) (solid blue curve)  and of its four-Gaussian approximation \eqref{4g} (dashed orange curve).  \subref{fig:subfigag} When plotted as a function of the signal position, the four-Gaussian gives a reasonable estimate of the general wavefunction for photons with the same wavelength without discriminating a specific photon's position. 
The general wavefunction requires wavelength degenerated photons when the pump pulse duration is long \subref{fig:subfigbg}, and allows a certain bandwidth for shorter pump pulse duration \subref{fig:subfigcg}. Results obtained for a SPDC type-I photons exiting a LiIO3 crystal of length $L=5\,\mathrm{mm}$, pump wavelength $\lambda_p=400\,\mathrm{nm}$ and beam waist $\sigma_p=28\,\mathrm{\mu m}$.}
\label{dgplots}
\end{figure}

\begin{center}
\textbf{\footnotesize Type-II SPDC}
\end{center}
\vspace{0.3cm}
\label{app:c}

\begin{figure}[ht!]
\centering
\subfloat[(a)]{\includegraphics[width=123pt]{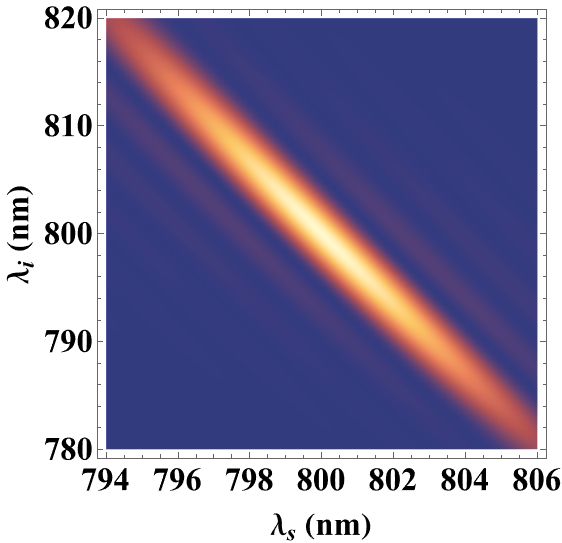}\label{fig:sinctii}}
\subfloat[(b)]{\includegraphics[width=123pt]{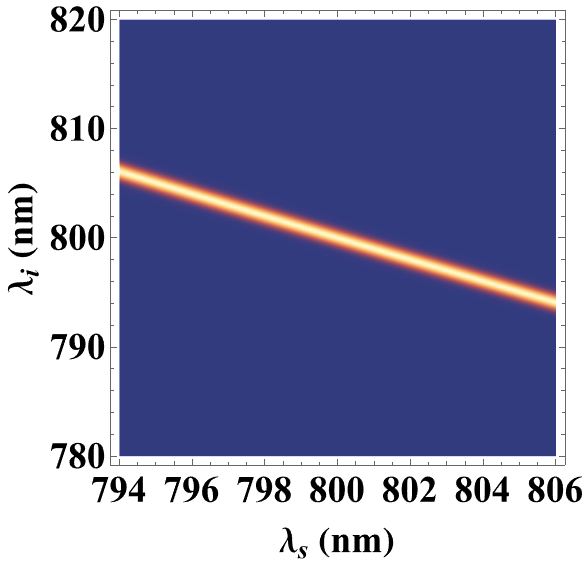}\label{fig:gausstii}}\\
\subfloat[(c)]{\includegraphics[width=123pt]{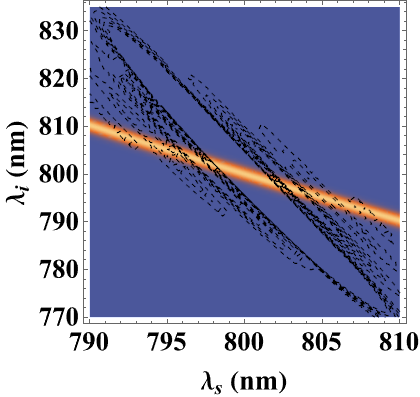}\label{fig:gausstii}}
\caption[Optional caption for list of figures]{Joint spectral amplitude (JSA) profiles \subref{fig:sinctii} obtained from the sinc-based wavefunction (\ref{generalII})-(\ref{sincsII}), and \subref{fig:gausstii} from the four-Gaussian approximation \eqref{4f2} as a function of the signal and idler wavelengths. We overlay the JSA from the Gaussian model as a density plot and the JSA from the sinc-based model as contour lines. We consider a BBO crystal with a  phase-matching angle $\theta \simeq 42.4^\circ$, whose length is $L=5\,\mathrm{mm}$, pump wavelength $\lambda_p = 400\,\mathrm{nm}$, pump beam waist $w_p = 28\,\mathrm{\mu m}$, pump pulse duration $\tau=50\,\mathrm{ps}$, signal and idler spatial coordinate is fixed at $x_s=-x_i=1\,\mathrm{\mu m}$, group velocities $v_{g, p}=c/1.708$, $v_{g, s}=c/1.626$ and $v_{g, i}=c/1.684$, group velocity dispersion $\beta_p=180\,\mathrm{fs}^2/\,\mathrm{mm}$, $\beta_s=61.7\,\mathrm{fs}^2/\,\mathrm{mm}$ and $\beta_p=75.1\,\mathrm{fs}^2/\,\mathrm{mm}$ \cite{snlo2003}.}
\label{fig:jsa_tii}
\end{figure}

Defining the general wavefunction for the SPDC type-II as

\begin{equation}
    \begin{split}
         \Phi_{\text{II}}= & \mathrm{e}^{-w_p^2|\vec{q_s}+\vec{q_i}|^2} \mathrm{e}^{-{\frac{\tau^2}{8 \ln{2}}(\Omega_s+\Omega_i)^2}}    \sinc \left [\frac{\Delta k_z L}{2}\right], 
    \end{split}
    \label{generalII}
\end{equation}
and building upon the approach used for type-I SPDC, we now construct an approximate form of the biphoton wavefunction for type-II SPDC. Starting from the separable spatial and spectral phase-matching components 

\begin{equation}
    \begin{split}
         \Phi_{\text{II}} \approx & \mathrm{e}^{-w_p^2|\vec{q_s}+\vec{q_i}|^2} \mathrm{e}^{-{\frac{\tau^2}{8 \ln{2}}(\Omega_s+\Omega_i)^2}}\Theta(q_s, q_i)  \Tilde{\Theta}(\Omega_s,\Omega_i),  
    \end{split}
    \label{b6}
\end{equation}
with
\begin{equation}
\begin{split}
\Theta( \vec{q_s}, \vec{q_i})=&\mathrm{e}^{-w_p^2|\vec{q_s}+\vec{q_i}|^2}  \sinc \left[\frac{L}{4k_p} |\vec{q_s}-\vec{q_i}|^2\right],\nonumber
\end{split}
\label{b7}
\end{equation}   
and
\begin{equation}
\begin{split}
\Tilde{\Theta}(\Omega_s,\Omega_i)
=
\sinc\!\Biggl[\frac{L}{2}\Biggl(&  -\frac{(\sqrt{\beta_{s}}\Omega_s-\sqrt{\beta_{i}}\Omega_i)^2}{4}\\ 
    &+\frac{1}{v_g^p}(\Omega_s+\Omega_i)-\frac{1}{v_g^{s}}\Omega_s-\frac{1}{v_g^{i}}\Omega_i\Biggr)\Biggr].
\end{split}
\label{b8}
\end{equation}

\begin{equation}
\begin{split}
   \psi (x_s, x_i, \Omega_s, \Omega_i)= &\, \mathrm{e}^{-\frac{(x_s+x_i)^2}{16\sigma_p^2}}\, \mathrm{e}^{-\frac{(x_s-x_i)^2}{16\sigma_x^2}} \\
   &\times \mathrm{e}^{-\alpha\frac{(\Omega_s+\Omega_i)^2}{2a^2(\tau)}}\, \mathrm{e}^{-\frac{(\sqrt{\beta_{s}}\Omega_s - \sqrt{\beta_{i}}\Omega_i)^2}{2b^2(\tau)}}.
\end{split}
\label{4f2}
\end{equation}
$a(\tau)$ and $b(\tau)$ are given by \eqref{ab}, with $B=\frac{\omega_0 c }{4}$. As for the type-I approximation, in type-II \( \alpha = 1 \) for long pulses ($\tau=50\,\mathrm{ps}$) and \( \alpha = 0.4 \) for short pulses ($\tau=50\,\mathrm{fs}$), ensuring that the absolute square of this approximation closely matches that of the sinc-based wavefunction across arbitrary pulse duration regimes. Refer to Fig.~\ref{dgplotsII} for a graphical comparison between the squared modulus of the Gaussian approximation and the sinc-based wavefunction for long and short pump pulse duration. A comparison of the Joint Spectral Amplitude (JSA) obtained using this four-Gaussian model and the sinc-based wavefunction for the SPDC type-II is depicted in \figref{fig:jsa_tii}. 


\captionsetup[subfloat]{justification=raggedright,singlelinecheck=false,position=top, labelformat=empty}

\def\ccc{115pt}
\begin{figure}[ht]
\centering
    \subfloat[]
    [
    (a) $\tau=50\,\mathrm{ps}$, $\lambda_i=800\,\mathrm{nm}$\newline
     \phantom{=}\,\phantom{=}\,$x_s=-x_i=1\,\mathrm{\mu m}$ 
    ]
    {\includegraphics[width=\ccc]{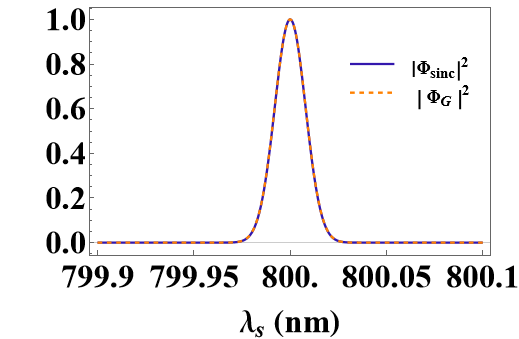}
    \label{fig:subfigbII}
    }
    \subfloat[][
    (b) $\tau=50\,\mathrm{fs}$,  $\lambda_i=800\,\mathrm{nm}$ \newline
     \phantom{=}\,\phantom{=}\,$x_s=x_i=1\,\mathrm{\mu m}$ 
    ]{
    \includegraphics[width=\ccc]{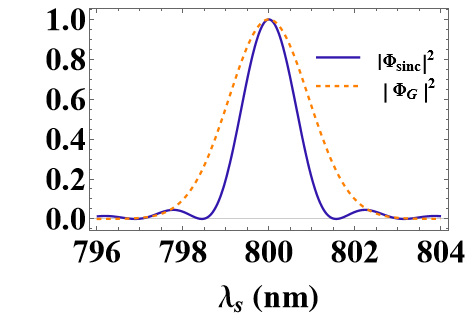}
    \label{fig:subfigcII}
    }
\caption[Optional caption for list of figures]{Squared modulus of the general type-II wavefunction Eq. (\ref{generalII})-(\ref{sincsII}) (solid blue curve) with its four-Gaussian approximation ~\eqref{4f2} (dashed orange curve) as a function of the signal wavelength.
Different pump pulse regimes are analyzed, long ($\tau = 50\,\mathrm{ps}$) \subref{fig:subfigbII} and short pulse duration ($\tau = 50\,\mathrm{fs}$) \subref{fig:subfigcII}, all the other parameters are the same as the ones used in \figref{fig:jsa_tii}. As in the type-I case, a longer pulse constrains the spectral bandwidth of the twin-photons, while a shorter pulse broadens it. The type-II spectrum is slightly wider than type-I, a feature well captured by the four-Gaussian approximation with the proper $\alpha$ adjustment in \eqref{4f2}.}
\label{dgplotsII}
\end{figure}

In Fig.~\ref{dgplotsII}, we compare the squared modulus of the general type-II wavefunction Eq. (\ref{generalII})-(\ref{sincsII}) (solid blue curve) with its four-Gaussian approximation ~\eqref{4f2} (dashed orange curve) as a function of the signal wavelength.
 Panels \subref{fig:subfigbII} and \subref{fig:subfigcII} illustrate the cases of long (\( \tau= 50\,\mathrm{ps} \)) and short (\( \tau= 50\,\mathrm{fs} \)) pump pulses, respectively. We observe a similar behavior obtained for type-I: a long pump pulse yields a narrow spectral bandwidth, and a wider bandwidth for short pulse. Additionally, we note that the spectral bandwidth for type-II is slightly broader than for type-I. This broader spectrum is well reproduced by the approximation with an appropriate choice of \( \beta \) in \eqref{4f2}.

For wavelength-degenerate photons, the pump pulse durations will not affect the spatial profile of the photons, as we are using parameters describing a region where the two degrees of freedom are decoupled. Since the resulting behavior is similar to what is shown in Fig.~\ref{dgplots}\subref{fig:subfigag} for type-I, we do not repeat it here. Compared to the type-I case, the main difference in the approximate model lies in the spectral profile, due to the asymmetric GVD contributions in type-II.

\begin{figure}[ht]
\subfloat[]
    [(a) GVD$_s=1$, GVD$_i=2$]
{\includegraphics[width=0.23\textwidth]{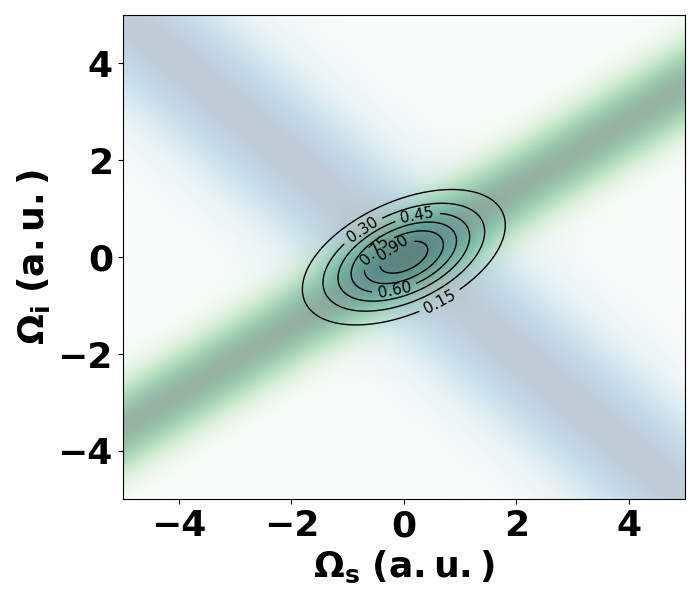}
\label{fig:jsa}}
\subfloat[]
    [(b) GVD$_s=1$, GVD$_i=0.25$]
{\includegraphics[width=0.23\textwidth]{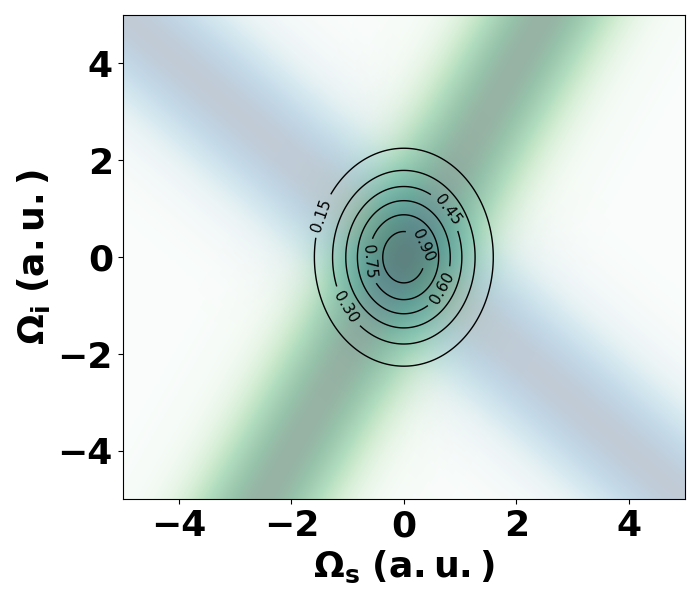}
\label{fig:jsb}}\\
\subfloat[]
    [(c) GVD$_s=0$, GVD$_i=5$]
{\includegraphics[width=0.23\textwidth]{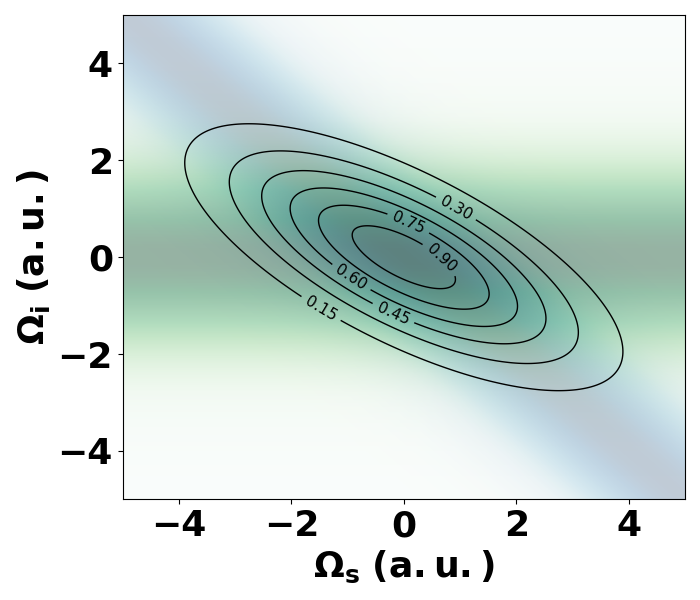}
\label{fig:jsc}}
\caption[Optional caption for list of figures]{JSA profiles generated using the proposed four-Gaussian model for type-II SPDC, which explicitly incorporates the signal and idler group velocity dispersion. To illustrate the qualitative effect of the signal and idler GVD on the JSA we use here arbitrary parameters, which do not correspond to experimental values. The JSA is obtained from the interplay between the two effective spectral Gaussians constraints; the sum-frequency and GVD-weighted coordinate $\sqrt{\beta_s} \Omega_s- \sqrt{\beta_i}\Omega_i$  Gaussian and the GVD-weighted. These results illustrate how the GVD asymmetry in the type-II model influence the orientation and ellipticity of the distribution.}
\label{fig:jsa_gvd_effects}
\end{figure}





Taking only the spectral counterpart of \eqref{4f2}, let us qualitatively investigate the effect of the signal and idler GVD on the JSA. We fix the parameters at $\beta=0.5$, $a(\tau)=b(\tau)=1$, while varying GVD$_s$ and GVD$_i$.
As shown in the overlaid plots in \figref{fig:jsa_gvd_effects}, for an intermediate pump duration, the orientation and ellipticity of the JSA can be affected by the signal and idler GVD coefficients. In panels \subref{fig:jsa} and \subref{fig:jsb}, the Joint Spectral Amplitude (JSA) is visibly elliptical, while in panel \subref{fig:jsc}, the JSA to become nearly circular.

The asymmetry in dispersion (arising from the phase-matching configuration in type-II SPDC, where signal and idler photons are orthogonally polarized propagating differently in a birefringent medium) plays a central role in shaping the joint spectral amplitude. Our four-Gaussian model captures this effect by explicitly incorporating distinct GVD terms for the signal and idler into the spectral Gaussian that represents the phase-matching function. This level of control offers a tunable framework for tailoring group-velocity conditions in high-purity photon applications, such as heralded single-photon sources and quantum-interference schemes.

\newpage

\section{LiIO$_3$ $\mathrm{vs}$ BBO-SPDC Type-I}
\label{app:c}

\begin{figure}[ht!]
\centering
\subfloat[\hspace{4cm} LiIO$_3$\\  \vspace{0.2cm} (a)]{\includegraphics[width=120pt]{accepted/cpm.png}\label{fig:subfigaliio}}
\subfloat[\hspace{0.5cm}(b)]{\includegraphics[width=120pt]{accepted/pma.png}\label{fig:subfigbliio}}\\[-5pt]
\subfloat[\hspace{4cm} BBO\\(c) ]{\includegraphics[width=120pt]{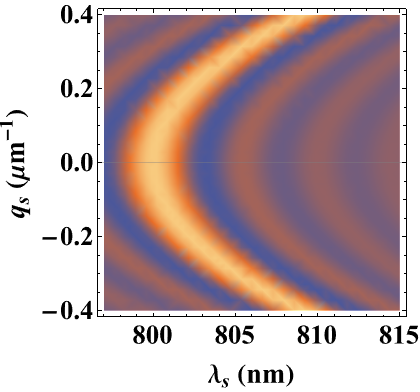}}\label{fig:subfigsincpm}
\subfloat[ \hspace{0.5cm}(d)]{\includegraphics[width=120pt]{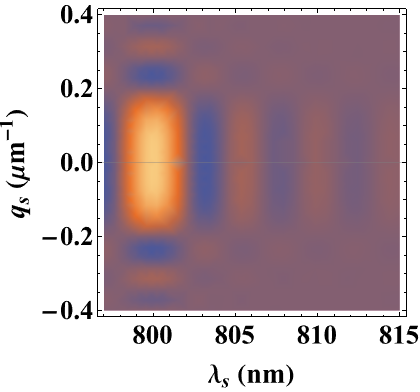}\label{fig:subfigapp}}\\
\subfloat[ (e)]{\includegraphics[width=120pt]{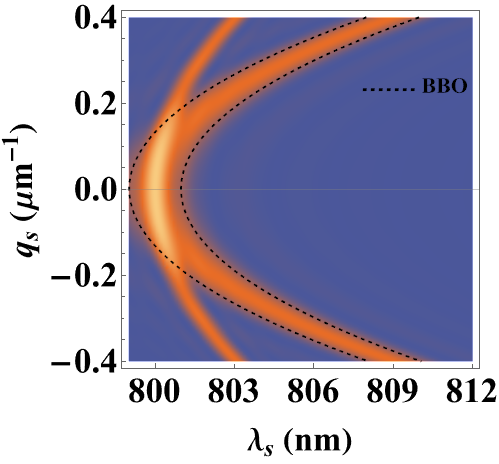}}\label{fig:subfigsuperpsinc}
\subfloat[(f)]{\includegraphics[width=120pt]{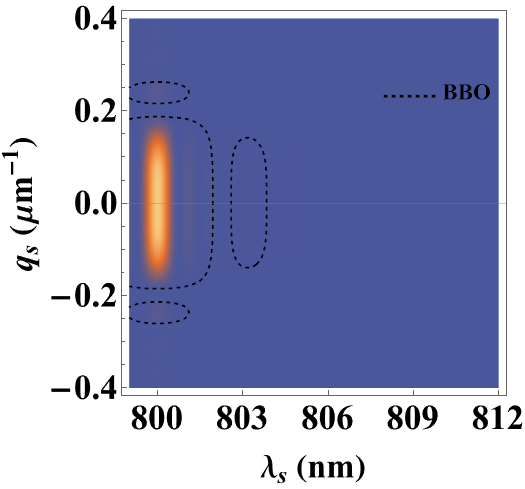}\label{fig:subfigsuperpgauss}}
\caption[]{PMF in the general and approximate models of type-I SPDC  for two types of nonlinear crystal: (Top) LiIO$_3$, with a cut angle $\theta \simeq 42.5^\circ$, and (Bottom) BBO, with  $\theta \simeq 29.2^\circ$.
 We compare the spatially and spectrally coupled sinc phase-matching function (PMF), computed using ~\eqref{sincsum} and shown in panels \hyperref[fig:subfigaliio]{(a)} and \hyperref[fig:subfigsincpm]{(c)}, with the double-sinc approximation of the phase-matching function given by ~\eqref{sincmult2} and shown in panels \hyperref[fig:subfigbliio]{(b)} and \hyperref[fig:subfigapp]{(d)}. 
   The PMFs are shown as a function of the signal wavelength $\lambda_s$ and transverse momentum $q_s$ (restricted to the $x$-direction), with fixed idler wavelength $\lambda_i = 800~\mathrm{nm}$ and transverse momentum $q_i = 0.01~\mu\mathrm{m}^{-1}$. For easy of qualitative comparison, we plot the sum of the absolute squares of the sinc-based PMF ($|\sinc_{\text{LiIO}_3}( q, \Omega)|^2$+$|\sinc_{\text{BBO}}( q, \Omega)|^2$) for both nonlinear materials in \hyperref[fig:subfigsuperpsinc]{(e)}. In \hyperref[fig:subfigsuperpgauss]{(f)}, the absolute squares of the corresponding approximate PMFs LiIO$_3$ and BBO are shown  separately as a density and contour plots, to facilitate their distinction since the two distributions overlap in certain regions.}
    \label{sincBBO}
\end{figure}

We now apply the toolkit for type-I SPDC we developed above to a type of crystal that is more widely used in modern applications: beta-barium borate (BBO) \cite{BBO}. We begin by comparing the phase-matching function  obtained for LiIO$_3$ and BBO for the same set of parameters, as shown in \figref{sincBBO}.

 For each crystal, we consider the spatial-spectral correlated sinc function computed through \eqref{sincsum}, and its correlation-free approximation given by two separate sinc functions ~\eqref{sincmult2}-(\ref{sincmult}). Compared to LiIO$_3$ \hyperref[fig:subfigaliio]{(a)}, BBO exhibits a broader phase-matching band whose wavelength dependence changes more gradually as the spatial coordinate varies \hyperref[fig:subfigsincpm]{(c)}.
 
 This behavior indicates a weaker spatial-spectral correlation over a broader bandwidth, as reflected by a larger region over which the PMF can be approximated by treating the spatial and spectral degrees of freedom as independent (compare \hyperref[fig:subfigbliio]{(b)} and \hyperref[fig:subfigapp]{(d)}).


\captionsetup[subfloat]{justification=raggedright,singlelinecheck=false,position=top, labelformat=empty}

\clearpage


In \figref{sincGbbo}, we compare the PMF and resulting wavefunctions for both crystals as function of only one variable at a time. The central peak of the sinc PMF as a function of the signal wavelength is considerably broader for BBO than for LiIO$_3$  \subref{fig:subfigsinclamb2}. The side lobes of the BBO PMF also extend over a wider signal wavelength range. BBO has smaller group velocity mismatch between the pump and the twin-photons than LiIo$_3$, as well as lower group velocity dispersion \cite{snlo2003}. In particular, as a consequence of the larger temporal walk-off for LiIO$_3$, the phase mismatch varies more rapidly with wavelength, resulting in a narrower spectral PMF.

Meanwhile, only a slight difference is observed between the sinc PMFs as a function of the transverse momentum \subref{fig:subfigsinclamb3}. Within the paraxial approximation used here the material dependence comes form the spatial phase-matching governed by the transverse-momentum quadratic terms only, which are inversely proportional to $k_p=\frac{2 \pi}{\lambda_p} n_{e,eff}(\lambda_p,\theta)$. Since the relevant wavevectors of BBO and LiIO$_3$ are similar under  the considered conditions, the transverse-momentum range permitted by their PMFs are also similar. 

Lastly, we analyze how well the spatial-spectral correlation-free Gaussian-based model approximates the sinc-based SPDC wavefunction. As already discussed, additionally to the phase-matching, the pump beam has a considerable influence on the properties of the twin-photons. As for LiIO$_3$, the emission is restricted to an almost degenerate-wavelength configuration for BBO in the long pump pulse duration regime \subref{fig:subfig4gshort2}. For shorter pump pulses, BBO supports an even broader bandwidth than LiIO$_3$, in agreement with discussion above. As in the case of LiIO$_3$, the four-Gaussian model \eqref{4g} provides a reasonable approximation to the general sinc-based model \eqref{eq:phi-sinc}-(\ref{sincsum}), in which the spatial and spectral degrees of freedom are coupled, for BBO in both pump duration regimes.

\def\ccc{115pt}
\begin{figure}[ht]
\centering
    \subfloat[][
    (a) $\lambda_i=800\,\mathrm{nm}$ \newline
     \phantom{=}\,\phantom{=}\,$q_s=-q_i=0.01\,\mathrm{\mu m^{-1}}$ 
    ]{
    \includegraphics[width=\ccc]{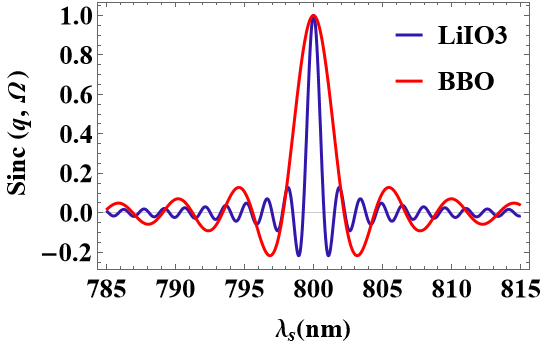}
    \label{fig:subfigsinclamb2}
    }   
    \subfloat[][
     (b) $\lambda_s=\lambda_i=800\,\mathrm{nm}$\newline
     \phantom{=}\,\phantom{=}\,$q_i=0.01\,\mathrm{\mu m^{-1}}$ 
    ]{
    \includegraphics[width=\ccc]{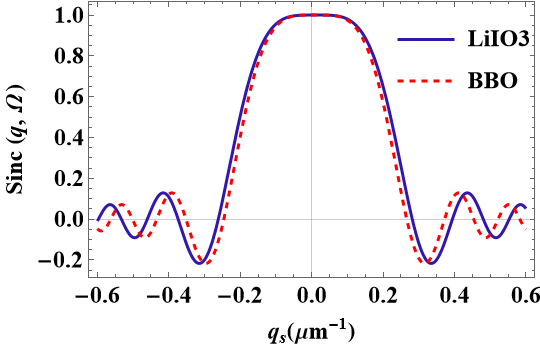}    \label{fig:subfigsinclamb3}}
\\
        \subfloat[][
    (c) $\tau=50\,\mathrm{ps}$,  $\lambda_i=800\,\mathrm{nm}$ \newline
     \phantom{=}\,\phantom{=}\,$q_s=-q_i=1\,\mathrm{\mu m^{-1}}$  
    ]{
    \includegraphics[width=\ccc]{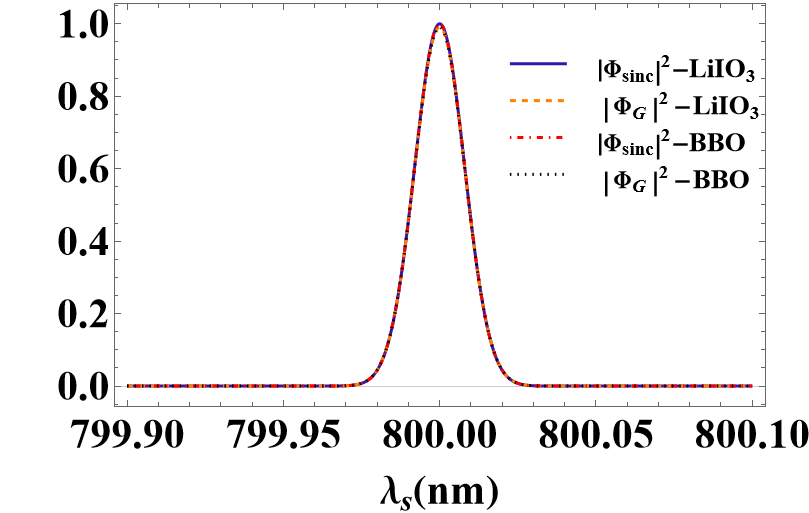}
    \label{fig:subfig4glong2}
    }
    \subfloat[][
    (d) $\tau=50\,\mathrm{fs}$,  $\lambda_i=800\,\mathrm{nm}$ \newline
     \phantom{=}\,\phantom{=}\,$q_s=-q_i=1\,\mathrm{\mu m^{-1}}$  
    ]{
    \includegraphics[width=\ccc]{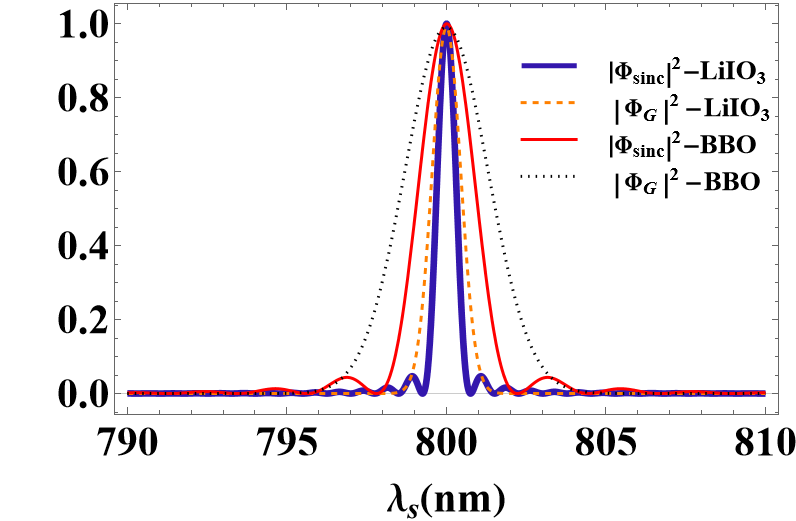}
    \label{fig:subfig4gshort2}
    }

\caption[Optional caption for list of figures]{We compare in more details a sinc PMF given through the phase mismatch  \eqref{sincsum} for LiIO$_3$ and BBO crystals. \subref{fig:subfigsinclamb2} shows the PMF as a function of the signal transverse momentum $q_s$, while \subref{fig:subfigsinclamb3} shows it as a function of the signal wavelength $\lambda_s$. The squared modulus of the wavefunction for a long ($\tau=50\,\mathrm{ps}$) and short ($\tau=50\,\mathrm{fs}$) pump pulse, as a function of $\lambda_S$, is shown in \subref{fig:subfig4glong2} and \subref{fig:subfig4gshort2}, respectively. For each nonlinear material, we consider two wavefunctions: (1) the general model expressed in terms of the sinc function, which retains the spatial-spectral correlation, $|\Phi_{sinc}|^2$, as obtained from \eqref{eq:phi-sinc}-(\ref{sincsum}); and (2) the correlation-free four-Gaussian approximation given by \eqref{4g}. We use the same set of parameters for both crystal types, except for those that are material dependent: $k_p=\frac{2 \pi}{\lambda_p} n_{e,eff}(\lambda_p,\theta)$, with $\theta_{LiIO_3} \simeq 42.5^\circ$ and $\theta_{BBO}= \simeq 29.2^\circ$, and in particular from \eqref{AB}, $\text{A}=0.17$ and $\text{B}=0.069$ for LiIO$_3$, whereas $\text{A}=0.058$ and $\text{B}=0.026$ for BBO.}
\label{sincGbbo}
\end{figure}

\clearpage

\bibliography{sequoia}

\end{document}